\documentclass[12pt]{article}
\usepackage{amsmath,amssymb,epsfig}
\makeatletter \@addtoreset{equation}{section} \makeatother
\renewcommand{\theequation}{\thesection.\arabic{equation}}
\newcommand{\ba}{\begin{array}}
\newcommand{\ea}{\end{array}}
\newcommand{\beq}{\begin{equation}}
\newcommand{\eeq}{\end{equation}}
\newcommand{\bea}{\begin{eqnarray}}
\newcommand{\eea}{\end{eqnarray}}

\def\bce{\begin{center}}
\def\ece{\end{center}}

\def\nonu{\nonumber}

\def\be{\beta}

\def\eps6{{\displaystyle \mathop{\epsilon}^{6}}{}}

\def\nab6{{\displaystyle \mathop{\nabla}^{6}}{}}

\def\ft#1#2{{\textstyle{\frac{\scriptstyle #1}{\scriptstyle #2}}}}
\def\fft#1#2{\frac{#1}{#2}}
\def\0{{\sst{(0)}}}
\def\1{{\sst{(1)}}}
\def\2{{\sst{(2)}}}
\def\3{{\sst{(3)}}}
\def\4{{\sst{(4)}}}
\def\5{{\sst{(5)}}}
\def\6{{\sst{(6)}}}
\def\7{{\sst{(7)}}}
\def\8{{\sst{(8)}}}

\def\nnn{\nonumber}

\def\ba{\begin{array}}
\def\ea{\end{array}}
\def\beq{\begin{equation}}
\def\eeq{\end{equation}}
\def\be{\begin{equation}}
\def\ee{\end{equation}}

\def\eps{\epsilon}

\def\ba{\begin{array}}
\def\ea{\end{array}}
\def\beq{\begin{equation}}
\def\eeq{\end{equation}}
\def\be{\begin{equation}}
\def\ee{\end{equation}}

\def\eps{\epsilon}

\newcommand{\bean}{\begin{eqnarray*}}
\newcommand{\eean}{\end{eqnarray*}}

\begin{document}
\thispagestyle{empty} \addtocounter{page}{-1}
\begin{flushright}
{\tt hep-th/0605012}\\
\end{flushright}

\vspace*{1.3cm}

\centerline{ \Large \bf Quark-Monopole
Potentials from}
\vspace{.3cm} 
\centerline{ \Large \bf  Supersymmetric $SL(3,R)$ Deformed IIB Supergravity} 
\vspace*{1.5cm}
\centerline{{\bf Changhyun Ahn}} 
\vspace*{1.0cm} 
\centerline{\it
Department of Physics, Kyungpook National University, Taegu
702-701, Korea} 
\vspace*{0.8cm} 
\centerline{\tt ahn@knu.ac.kr} 
\vskip2cm

\centerline{\bf Abstract}
\vspace*{0.5cm}

We recompute the quark-monopole potential 
from supersymmetric $SL(3,R)$ deformation of IIB supergravity 
background dual to 
deformed Coulomb branch flow of the ${\cal N}=4 $ super Yang-Mills theory.  
The marginal deformations
strengthen
the Coulombic attraction between quarks and monopoles. 

\baselineskip=18pt
\newpage
\renewcommand{\theequation}
{\arabic{section}\mbox{.}\arabic{equation}}

\section{Introduction}


The $SL(3,R)$ symmetry of type IIB supergravity on a two-torus
has been applied to finding new gravity solutions which 
correspond to marginal deformation of ${\cal N}=4$ super Yang-Mills
theory \cite{LM}.
This method can be used to any solution that has an isometry group
which contains $U(1) \times U(1)$. If there exists an extra 
$U(1)_R$-symmetry in addition to this symmetry, then the 
deformed solution preserves ${\cal N}=1$ supersymmetry.

In particular, the gravity description of deformed Coulomb branch 
renormalization group flow 
of ${\cal N}=4$ super Yang-Mills theory with $SO(2)^3$ global symmetry   
has been discussed in \cite{AV}. The UV limit of 
the dual gauge theory is the Leigh-Strassler deformation \cite{LS} of
${\cal N}=4$ super Yang-Mills theory. This global symmetry corresponds to
$U(1) \times U(1)$ global symmetry generated by two angles of two torus
and 
$U(1)_R$ symmetry generated by remaining angle of internal space.

In \cite{AV1}, the Coulombic potential between quark and anti-quark
has been 
reconsidered for supersymmetric $SL(3,R)$ deformed type IIB theory
dual to the deformed Coulomb branch flow of ${\cal N}=4$
super Yang-Mills theory with $SO(4) \times SO(2)$ global symmetry.
Two  of three parameters determining the shape of D3-branes become 
identical to each other for purely radial string configuration leading to
symmetry enhancement $SO(2)^3 \rightarrow SO(4) \times SO(2)$.
One of the main results is that for certain part of the moduli space,
the $\sigma$ deformations induce a transition from Coulombic attraction 
between quarks and anti-quarks to linear confinement.
For the undeformed solution where $\sigma=0$, 
there was no confining behavior but as the 
distance becomes larger, this transition to a regime of linear confinement
occurs where the scale of confinement increases with $\sigma$.     

Now it is natural to compute the potential between two magnetic monopoles
in terms of D-string worldsheet and the result for conformal theory
will be the same as 
the one in \cite{RY,Malda} but with $g_{YM} \rightarrow 4\pi/g_{YM}$ or
$g \rightarrow 1/g$. One can also
compute the interaction between a quark and a magnetic monopole. In this case,
the fundamental string ending on a quark will attach to the D-string
ending on a magnetic monopole and they will connect to form a $(1,1)$ 
string
which will go into the horizon \cite{Minahan}.  
In particular, when the string coupling $g$ is small, 
the D-string is very rigid and the 
fundamental string will end on D-string perpendicularly. 
Then the solution 
for the fundamental string will be half of the solution for two fundamental
strings and leads to $1/4$ in the potential \cite{Minahan} in 
$AdS_5 \times S^5$ background. 

In this paper, we extend the result of \cite{AV1} to the case of 
a massive quark and monopole by replacing an anti-quark with a monopole 
or generalize the result of \cite{Minahan}
corresponding to $AdS_5 \times S^5$ background of type IIB theory
to the case of computation in the supersymmetric $SL(3,R)$ deformed type IIB
theory by considering more general background. We will see 
the marginal deformations  of ${\cal N}=4$ super Yang-Mills theory  
strengthen
the Coulombic attraction between quarks and monopoles. 
On the gauge theory side, similar discussion can be found in \cite{Dorey}.

\section{Quark-monopole potentials revisited}

We study the effect of $\sigma$ deformation on the quark and monopole 
potential for (non)conformal field theories. The quark and monopole 
potential has been studied for ${\cal N}=4$ super Yang-Mills theory in the 
large $N$ limit \cite{Minahan} of $AdS_5 \times S^5$ type IIB background.
Now we extend this background to 
supersymmetric $SL(3,R)$ deformed type IIB theory
dual to the deformed Coulomb branch flow of ${\cal N}=4$
super Yang-Mills theory.
The UV limit of undeformed($\sigma=0$) theory  
asymtotes to $AdS_5 \times S^5$ where the conformal symmetry is
regained. 

The supergravity dual of the deformed Coulomb branch flow in the string
frame is \cite{AV,AV1}
\bea
ds^2 &=& \alpha^{\prime}\sqrt{Hf}\, R^2
\left( 
\fft{r^2}{fR^4}\,
dx_{\mu}^2 +\fft{dr^2}{f r^2 L_1 L_2 L_3} 
+R^2 \, ds_{\tilde{S}^5}^2
\right)
\label{metric}
\eea
where the length scale is $R^4= g_{YM}^2 N$, the metric of deformed 
five-sphere $\tilde{S}^5$
depends on the internal coordinates $\alpha, \theta$ 
and the various functions are
given by 
\bea
 L_i = 1+\fft{\ell_i^2}{r^2}\,, 
\qquad 
H=1+\hat\sigma^2 \fft{f}{gh} \, s_{\alpha}^2\,, \qquad \hat\sigma \equiv
\sigma R^2/2.
\nonu 
\eea
The three parameters $\ell_i$ specify the ellipsoidal shape of 
D3-brane distribution \cite{KLT}. We assume that the classical supergravity
description is valid with the same spirit of \cite{AV1}. 
The explicit form for functions $f,g$ and $h$
is given in \cite{AV,AV1} but we do not need them in this paper. For
the purely radially oriented string configuration, they have simple 
expression. Moreover, $H$ and $f$ can be written in terms of 
$L_i$'s. See the footnote 1.
The metric of deformed five-sphere depends on the modulus of
$\beta=\gamma - \tau_s \sigma$ where $\gamma$ and $\sigma$ are real
deformation parameters and $\tau_s$ is a complex structure.

Now  a probe D3-brane has been taken at large distance($r=\infty$) 
and let us consider the behavior of string configurations ending on this 
brane. The $N$ D3-branes are located at $r=0$.
To find a static configuration we set $\tau=T$ and $\sigma=x$
where $x$ is a direction along D3-branes.
Then string action can be simplified to \cite{AV1}
\bea 
S=\fft{T}{2\pi} \int dx \sqrt{H} \,
\sqrt{\fft{r^4}{fR^4}+\fft{r^{\prime 2}}{fL_1L_2L_3} 
} 
\label{action}
\eea
where $\prime$ denotes a derivative with respect to $x$ and $T$ denotes
the time interval. We put the the solution to the equations of motion 
:$\alpha=\pi/2, \theta=\pi/4$ and $\ell_2=\ell_3$.
When there are no D3-brane distributions($f=1=L_i$) 
and there are no $\sigma$ 
deformations($H=1$), then the above geometry is exactly the same as 
the conformal case of ${\cal N}=4$ super Yang-Mills theory 
\cite{Malda,RY}. For the deformed solution with no D3-brane distributions,
as in the case of quark-anti quark potential, the conformal factor $H$ is
a constant($H = 1 + \hat\sigma^2$). Then the Wilson loop computation implies
that the marginal deformations of ${\cal N}=4$ super Yang-Mills theory
enhance the Colombic attraction.  

There exist a heavy quark at $x=0$ and a heavy monopole at $x=L$ 
\cite{Minahan}.
These transform under the fundamental representation of $SU(N)$.
In the $(x,r)$-plane, a fundamental string(or $(1,0)$ string) 
is attached to the D3-brane
at $(0,\infty)$ and a D-string(or $(0,1)$ string) is attached to 
D3-brane at $(L,\infty)$. Moreover, 
there should be another $(1,1)$ string attached to the other strings
at $(\Delta L, r_0)$ and the other end of this $(1,1)$ string is 
attached to one of the D3-branes at $(\Delta L, 0)$.   
The detailed configuration was given in \cite{Minahan} or similar
configuration where there exists a shift by $\Delta L$ in the $x$-axis
is given in Figure 1.

As for the $(1,0)$ and $(0,1)$ strings, 
the minimization of the action for each string 
\footnote{In the below,
we use the reduced relations $H =1+\hat\sigma^2 \frac{L_2}{L_1}$ and $f=\frac{1}
{L_1 L_2}$ for purely radial string configuration all the time.}
satisfies  
\be 
\sqrt{\fft{H}{f}} \,
\fft{r^4}{\sqrt{\fft{r^4}{R^4}+\fft{r^{\prime
2}}{L_1L_2^2}}}=\sqrt{\fft{H_i}{f_i}} \, r_i^2 R^2\,, \qquad 
i=1,2 
\label{mini}
\ee
where $i=1$ is for the fundamental string $(1,0)$ while 
$i=2$ is for the D-string $(0,1)$ and $H_i \equiv H(r_i)$ and 
$f_i \equiv f(r_i)$. The $r_i$'s are determined later.
Using these equations (\ref{mini}), one can write down 
$x$ in terms of $r$.
The solutions for the lengths of two strings are
\bea
 \Delta L & = & R^2 
\int_{r_0}^{\infty} \fft{dr}{r^2 \sqrt{L_1}L_2
\sqrt{\fft{f_1Hr^4}{fH_1r_1^4}-1}}\,,   \nonu 
\\
L-\Delta L & = & R^2 
\int_{r_0}^{\infty} \fft{dr}{r^2 \sqrt{L_1}L_2
\sqrt{\fft{f_2Hr^4}{fH_2r_2^4}-1}}. 
\label{L}
\eea
By adding these, one gets the length $L$ which is a function of $r_0,
r_1$ and $r_2$ as well as $\ell_i$ and $\hat\sigma$.

\begin{figure}[ht]
   \epsfxsize=4.5in \centerline{\epsffile{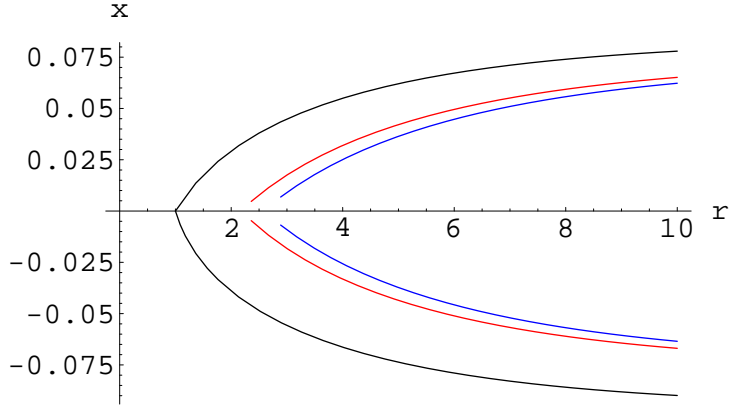}}
   \caption[FIG. \arabic{figure}.]{Various string configurations 
for $\ell_1=0$,
   $\ell_2=\ell_3=10$, $t=0.5$ and 
$\hat\sigma=1$ (black), $2$ (red) and $3$ (blue) using (\ref{L}).
   For a given $L$, 
the $\sigma$ deformations increase the energy scale probed by a string.
At $x=0, r=r_0$, $(1,1)$ string coming out of  this junction is
attached to one of $N$ D3-branes at the origin $x=0, r=0$ and $(1,0)$ and 
$(0,1)$ strings located at positive $x$ and negative $x$ respectively
are attached to the probe D3-brane at infinity($r=\infty$). 
Note that the $x=0$ of this figure is different from the one 
given in \cite{Minahan}.
 }
   \label{coulomb4}
\end{figure}

Figure 1 shows string configurations for  $\ell_1=0$,
   $\ell_2=\ell_3=10$ for a uniform distribution of 
D3-branes on a three-dimensional spherical shell. 
For a given a distance $L$ between quark and monopole, 
the $\sigma$ deformations
increase the energy scale probed by a string.
The behavior of $x=x(r)$ for the positive $x$ characterized by
$(1,0)$ string can be obtained 
from the first integral (\ref{L}) with the upper limit replaced by $r$.
On the other hand, the curve $x=x(r)$ with negative $x$ characterized by
$(0,1)$ string in Figure 1
can be obtained from the minus of the second integral (\ref{L}) with 
upper limit $r$. We also use the solutions for $r_1$ and $r_2$
which will be determined later by (\ref{r1r2}).
At fixed $r$, the distance between the horizontal axis and the curve 
$x=x(r)$ depends on the inverse of string coupling $t(\equiv 1/g)$. 
For $t < 1$($t=0.5$ for the case of
Figure 1), the distance between 
the horizontal axis to the curve with positive $x$ is smaller than 
the one to the curve with negative $x$.
For $t=1$, they are the same. For $t > 1$, the former is larger than
the latter.

The total regularized energy combined by the three kinds of 
string configurations  
is given by 
\bea 
E & = & \fft{1}{2 \pi } \left[ \int_{r_0}^{\infty} dr\left(
\sqrt{\fft{f_1}{H_1}}\, \fft{Hr^2}{fr_1^2 \sqrt{L_1} L_2
\sqrt{\fft{f_1Hr^4}{fH_1r_1^4}-1}}- \sqrt{1+\hat\sigma^2}\right)
- \int_{0}^{r_0} dr \sqrt{1+\hat\sigma^2} \right] \nonu
\\
&+ & \fft{ t}{2 \pi } (r_1 \rightarrow r_2)
+ \frac{1}{2\pi}   \sqrt{1+t^2} \int_{0}^{r_0} dr \sqrt{\frac{H}{L_2}} 
\label{E}
\eea
where the infinite contributions from the quark and monopole have been
subtracted in the first line and second line respectively. 
Note that the worldsheet action for any $(p,q)$ string has an
Euclideanized string worldsheet multiplied by a factor $\sqrt{p^2 + q^2
t^2}$ where $t$ is an inverse of string coupling $g$
\bea
t \equiv 1/g.
\nonu 
\eea
The last term of (\ref{E})
comes from the contribution of $(1,1)$ string 
where the main contribution arises from $r^{\prime}$ term inside the 
square root of the action (\ref{action}).

The length $L$ by adding the contributions (\ref{L}) and 
the energy $E$ from (\ref{E}) can be written in terms of 
elliptic integrals for four separate cases.
Note that the integrands of (\ref{L}) are the same as the case of
quark-anti quark potential \cite{AV1} except that $r_0$ is replaced by
$r_i$'s. So one can perform them without any difficulty. 
It turns out that elliptic integrals have more general arguments due
to the fact that lower limit $r_0$ is different from $r_i$. 
For the energy $E$, the
integrands from the contributions $(1,0)$ string and $(0,1)$ string
look similar to the ones in \cite{AV1} and can be computed
straightforwardly. The expressions are more complicated due to the 
presence of $r_i$ as well as $r_0$.
The contribution from $(1,1)$ string can be written as an elliptic
integral also.  

The first case \footnote{This case applies 
when either $r_i^2 \ge \ell_1^2-2\ell_2^2$ for
any $\hat\sigma$, or else $r_i^2 <\ell_1^2-2\ell_2^2$ and
$\hat\sigma^2<\ft{r_i^2+\ell_2^2}{\ell_1^2-2\ell_2^2-r_i^2}$.
The symbols $F,E,\Pi$ denote the elliptic integrals of the first,
second and third kind, respectively and $K$ denotes the complete
elliptic
integral of first kind.} 
is given by
\bea 
L &=& 2R^2 (1+\hat\sigma^2)\beta_1
\sqrt{(r_1^2+\ell_2^2)\alpha_1}\, [\Pi
(\alpha_0, \alpha_1,\sqrt{\alpha_2})-F(\alpha_0, \sqrt{\alpha_2})]
+ (r_1 \rightarrow r_2)\,,
\nnn\\
E &=& E_{1,0} + E_{0,1} +E_{1,1}
\label{EandL}
\eea
where the contributions to the energies
from $(1,0), (0,1)$ and $(1,1)$ strings are
given by
\bea
E_{1,0} & = &
\frac{\sqrt{\beta_2}}{2\pi} \left[ (1+\hat\sigma^2)(r_1^2+\ell_1^2)
\left( K(\sqrt{\alpha_2})-\frac{1}{1-\alpha_2} E(\sqrt{\alpha_2}) -
\Pi(\beta_0,1,\sqrt{\alpha_2})\right)
\right. \nonu \\
 &+& \left. (\ell_1^2 + \ell_2^2 \hat\sigma^2) 
F\left(\alpha_0,\sqrt{\alpha_2} \right)
 -   (1+\hat\sigma^2) \ell_1^2 \left(K(\sqrt{\alpha_2})-
F(\beta_0,\sqrt{\alpha_2}) \right) \right]\,, \nonu \\
E_{0,1} & =&  t E_{1,0}(r_1 \rightarrow r_2)\,, \nonu \\
E_{1,1}^{+} & = &
\frac{1}{2\pi}  \sqrt{1+\hat\sigma^2} \sqrt{1+t^2}
\left[ \frac{\sqrt{\ell_2^2-\ell_1^2}\,\hat\sigma^2}
{(1+\hat\sigma^2)} 
F\left(\mu, \frac{1}{\sqrt{1+\hat\sigma^2}}\right) \right.
\nonu \\
&-& \left.
\sqrt{\ell_2^2-\ell_1^2} \, E\left(\mu, 
\frac{1}{\sqrt{1+\hat\sigma^2}}\right)+ 
\sqrt{\frac{\alpha_1^0}{\beta_2^0(1+\hat\sigma^2)}}
-(r_0\rightarrow 0)\right]\,,
\nonu  \\
E_{1,1}^{-} & = & 
\frac{1}{2\pi}  \sqrt{1+\hat\sigma^2} \sqrt{1+t^2}
\nonu \\
&\times & \left[ -\frac{\sqrt{\ell_1^2-\ell_2^2}\,\hat\sigma}
{\sqrt{1+\hat\sigma^2}} 
E \left(\nu,\frac{\sqrt{1+\hat\sigma^2}}{\hat\sigma}\right)+ 
\sqrt{\frac{\alpha_1^0(r_0^2+\ell_1^2)}{\beta_2^0(1+\hat\sigma^2)(r_0^2+
\ell_2^2)}}
-(r_0\rightarrow 0)\right].
\label{Es}
\eea
Here the expression $(r_1 \rightarrow r_2)$ means 
that we simply change the quantity appeared previously by substituting 
$r_1$ with $r_2$. This applies to $L$ and $E_{0,1}$ above.
In the expression of $E_{1,1}$, the notation $(r_0 \rightarrow 0)$ holds
similarly. The $E^{+}_{1,1}$ is for the case $\ell_2 > \ell_1$
while $E_{1,1}^{-}$ is for the case $\ell_2 < \ell_1$.
The various parameters are given by
\bea 
\alpha_0 & \equiv & \sin^{-1} \sqrt{\frac{1}{1+\beta_2(r_0^2-r_1^2)(1+
\hat\sigma^2)}}\,, \qquad \beta_0 \equiv \sin^{-1} \sqrt{\frac{
r_0^2-r_1^2}{r_0^2+\ell_1^2}}\,,
\nonu \\
\alpha_i & \equiv & [r_1^2+\ell_i^2+\hat\sigma^2
(r_1^2+2\ell_2^2-\ell_{3-i}^2)]\beta_2\,,\qquad \beta_i^{-1}\equiv
2r_1^2+\ell_1^2+\ell_i^2+2\hat\sigma^2 (r_1^2+\ell_2^2)\,,
\nonu \\
\mu & \equiv & \sin^{-1} \sqrt{\frac{\beta_2^0(1+\hat\sigma^2)
(r_0^2+\ell_1^2)}{\alpha_1^0}}, \qquad
\nu  \equiv  \sin^{-1} \sqrt{\frac{\alpha_1^0}{\beta_2^0 
(1+\hat\sigma^2)
(r_0^2+\ell_2^2)}} 
\nonu
\eea
where $\alpha_1^0$ and $\beta_2^0$ are $\alpha_1$ and $\beta_2$
with the replacement $r_1 \rightarrow r_0$. 
One can easily check that the conformal case \cite{Minahan,Park} 
is recovered by putting $\ell_i=0$ and $\hat\sigma=0$ \footnote{
In order to see this, we need to use some properties between the elliptic
integrals \cite{AS}. 
When $\cos \alpha \tan \phi \tan \psi =1$, then 
there are two ``addition'' 
relations: $F(\phi,\alpha) + F(\psi,\alpha) =K(\alpha)$ 
and $E(\phi,\alpha)+E(\psi,\alpha)=E(\alpha) +\sin^2 
\alpha \sin \phi \sin \psi$.  Then our reduced expressions for $L$ and $E$
are exactly same as the ones in \cite{Park} where the zero-temperature limit
was given in terms of elliptic integrals rather than hypergeometric 
functions \cite{Minahan}.}. As we mentioned before, the energy of string
configurations for nonzero $\sigma$ is enhanced by $\sqrt{H}$ where 
$H = 1 +\hat\sigma^2$. The energy is $E = c \sqrt{H} /L$. The strength 
parameter $c$ is invariant under the S-duality transformation $g \rightarrow
1/g$. Since $L$ is also invariant under this transformation, $E$ is
invariant under the $g \rightarrow 1/g$ \cite{Minahan}.

Figure 2 is a parametric plot of the distance between the quark and
monopole $L$ versus the quark-monopole potential $E$ by using
(\ref{EandL}) and (\ref{Es}). For trajectories which are perpendicular to 
a uniform distribution of D3-branes on a particular five-dimensional
ellipsoid (all $\ell_i$ nonvanishing and $\ell_2=\ell_3$), the $\sigma$
deformations enhance the Coulombic force between the quark and
monopole. At asymptotically large distance, this force vanishes.
In Figure 2, this is shown for the case for $\ell_1=1$, $\ell_2=\ell_3=0.1$.
We set $R$ to unity for convenience because it has the effect of
rescaling $\hat\sigma$ and $L$. 
One can see similar behavior for the distribution of D3-branes
on a two dimensional disk($\ell_1 \neq 0, \ell_2=\ell_3=0$),
as in quark-anti quark potential \cite{AV1}. 

\begin{figure}[ht]
   \epsfxsize=4.0in \centerline{\epsffile{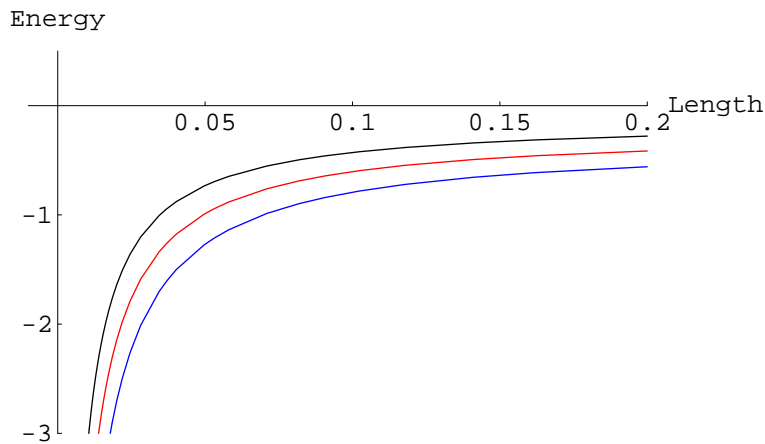}}
   \caption[FIG. \arabic{figure}.]{Quark-monopole potential $E=E(L)$ 
for $\ell_1=1$, $\ell_2=\ell_3=0.1$, $t=0.5$ and $\hat\sigma=1$
   (black), $1.5$ (red) and $2$ (blue) using a parametric plot for
(\ref{EandL}).
   In this part of the Coulomb branch, the $\sigma$ deformations simply
   enhance the Coulombic force. As we increase $t$, the whole curves
are shifted to the lower right direction. }
   \label{coulomb1}
\end{figure}

 

Assuming that the dominant contributions arise from the regions 
near $r=r_1$ for $\Delta L$ and near $r=r_2$ for $L-\Delta L$ in 
(\ref{L}), we find that
\bea
L & \approx & \frac{R^2}{r_1^2 \sqrt{L_1(r_1)} L_2(r_1)} I_1 +
\frac{R^2}{r_2^2 \sqrt{L_1(r_2)} L_2(r_2)} I_2
\nonu \\
E & \approx & \frac{1}{2\pi} \sqrt{\frac{H_1}{L_2(r_1)}} I_1 +
\frac{t}{2\pi} \sqrt{\frac{H_2}{L_2(r_2)}} I_2, 
\qquad
I_i  \equiv  \int_{r_0}^{\infty} \frac{dr}{
\sqrt{\frac{f_i H r^4}{f H_i r_i^4}-1}}.
\label{appro}
\eea
When $t=1$, then $r_1 =r_2$ from (\ref{r1r2}).
Since the $L$ and $E$ have common factor $I \equiv I_1=I_2$,
the energy $E$ is proportional to
the length $L$, as in quark-anti quark case \cite{AV1}.
In other words, the range of parameters($\alpha_2 \rightarrow 1, 
r_1, r_2 \rightarrow 0$ and $\ell_1 \rightarrow 0$) provides linear behavior
of quark-monopole. However, this range of parameters implies that
it takes an infinite amount of energy to separate quark and anti-quark.
Therefore, one cannot have a quark available to find the potential between
quark and monopole.  
When $t$ is not equal to 1$(r_1 \neq r_2)$, it is impossible to see any 
simple analytic expression between $E$ and $L$ because $r_1$ and $r_2$ are
complicated functions of $\ell_1, \ell_2, r_0$ and 
$\hat{\sigma}$ through (\ref{r1r2}). 

Following the procedure by \cite{Minahan}, we want to compute 
$r_i$  from the vanishing of net forces at the junction $r=r_0$ 
rather than differentiating the energy $E$ with respect to $r_i$.
For the $(1,0)$ string and $(0,1)$ string, the derivatives 
$r^{\prime}$ at $r=r_0$ can be obtained 
from (\ref{mini}) and the infinitesimal lengths squared along the
strings can be read off from the metric (\ref{metric}).
The $ds^2$ can be written in terms of $dx^2$.
Then according to \cite{Minahan} by recognizing that the integrand of
an action 
(\ref{action}) is equal to a tension $T_{p,q}$ multiplied by $ds$, 
the tensions of the strings at $r=r_0$ are
given by
\bea
T_{1,0} &= & \frac{1}{2\pi \sqrt{\alpha'} R} 
\left(\frac{H_0}{f_0}\right)^{1/4} r_0, \qquad
T_{0,1} =\frac{1}{2\pi \sqrt{\alpha'} R} 
t \left(\frac{H_0}{f_0}\right)^{1/4} r_0,
\nonu \\
T_{1,1} &=& 
\frac{1}{2\pi \sqrt{\alpha'} R} \sqrt{1+t^2}
\left(\frac{H_0}{f_0}\right)^{1/4} r_0.
\nonu
\eea
Of course, the conformal limit \cite{Minahan} is 
recovered since the extra factors
$\left(\frac{H_0}{f_0}\right)^{1/4}$
become 1 due to the fact that $H_0=f_0=1$.
From these, the vertical and horizontal components of forces 
exerted by each of the strings 
in the $(x,r)$-plane are set to zero \cite{DM,Sen}:
\bea
\frac{
\sqrt{\frac{H_1 }{f_1}} \, r_1^2
}{\sqrt{\frac{H_0  }{f_0}} \, r_0^2}  
\, T_{1,0} -
  \frac{
\sqrt{\frac{H_2 }{f_2}} \, r_2^2
}{\sqrt{\frac{H_0  }{f_0}} \, r_0^2}\,  T_{0,1} & = & 0, 
\nonu \\
\sqrt{1-\left(\frac{
\frac{H_1 }{f_1} \, r_1^4
}{\frac{H_0  }{f_0} \, r_0^4}\right)  }\, T_{1,0} +
 \sqrt{1-\left(\frac{
\frac{H_2 }{f_2} \, r_2^4
}{\frac{H_0  }{f_0} \, r_0^4}\right)  } \, T_{0,1}-T_{1,1} &= &0.
\label{equations}
\eea
Although the common extra factors
$\left(\frac{H_0}{f_0}\right)^{1/4}$
appearing the tension above do not change 
the relations between the forces, the slopes of $(1,0)$ string or
$(0,1)$ string at $r=r_0$ do depend on $H_i$ and $f_i$
where $i=0,1,2$. This will lead to 
more complicated expressions for $r_i$ where $i=1,2$.
By simplifying (\ref{equations}), the solution for these are
given by
\bea
\left(
\frac{\frac{H_0}{f_0} \, r_0^4}{\frac{H_1}{ f_1} \, r_1^4} \right)=
1 + \frac{1}{t^2}, \qquad
\left(
\frac{\frac{H_0}{ f_0} \, r_0^4}{\frac{H_2}{ f_2} \, r_2^4}\right) 
= 1+ t^2.
\label{r1r2}
\eea
Therefore, these allow us to write $r_1$ and $r_2$ in terms of $r_0,
\ell_i, \hat\sigma$
and $t$ by substituting $H_i$ and $f_i$ explicitly. 
It is easy to see the conformal limit \cite{Minahan} 
can be seen from the fact that
$H_i=1=f_i$ where $i=0,1,2$.
By substituting these expressions (\ref{r1r2}) with (\ref{L}) and (\ref{E}), 
all the previous plots on $x$ versus $r$ and $E$ versus $L$ 
are drawn for the values of $r_0, \ell_i, \hat\sigma$ and $t$. 
Note that there exists  a relation 
$r_1 \leftrightarrow r_2$
under the S-duality transformation $g \leftrightarrow 1/g$.

The second case \footnote{This case applies when $r_i^2<\ell_1^2-2\ell_2^2$ 
and
$\hat\sigma^2> \ft{r_i^2+\ell_2^2}{\ell_1^2-2\ell_2^2-r_i^2}$.}
for which $L$ and $E$ can be written in terms of 
elliptic integrals is
\bea 
L &=& \fft{R^2}{2(\ell_1^2-\ell_2^2)}
\sqrt{\fft{r_1^2+\ell_2^2}{2\beta_1
(1+\hat\sigma^2)(r_1^2+\ell_1^2)}} \,
[\Pi
\left(q_0,\ft{\ell_1^2-\ell_2^2}{r_1^2+\ell_1^2},
\sqrt{q}\right)-F(q_0,\sqrt{q})]\,
+ (r_1 \rightarrow r_2)\,,
\nnn\\
E &=& E_{1,0} + E_{0,1} +E_{1,1} 
\label{EandLsecond}
\eea
where the energy from $(1,0)$ string is given by 
\bea
E_{1,0} &=&
\frac{1}{2\pi  \sqrt{(1+
\hat\sigma^2)(r_1^2+\ell_1^2)}} \left[ (\ell_1^2+ \ell_2^2
\hat\sigma^2) 
F(q_0,\sqrt{q}) \right.
\nonu \\
&+&   
\frac{1}{\beta_2} (K(\sqrt{q})-\frac{1}{1-q} 
E(\sqrt{q}) -\Pi(p_0,1,\sqrt{q}))  
\nonu \\
&- & \left.  (1+\hat\sigma^2) \left( \frac{1}{\beta_2 (1+\hat\sigma^2)}-
r_1^2\right) (K(\sqrt{q}) -F(p_0,\sqrt{q}))  
\right],
\label{second}
\eea
the energy from $(0,1)$ string $E_{0,1}$ is equal to 
$t E_{1,0}(r_1 \rightarrow r_2)$
and finally the energy from $(1,1)$ string $E_{1,1}$ 
is the same as previous one (\ref{Es}).
The parameters in this case are defined by 
\bea 
q_0  \equiv  \sin^{-1} \sqrt{\frac{r_1^2+\ell_1^2}{r_0^2+\ell_1^2}},
\qquad
p_0 \equiv \sin^{-1} (\cos \alpha_0), \qquad 
q  \equiv  \fft{\hat\sigma^2
\ell_1^2-(1+2\hat\sigma^2)\ell_2^2-(1+\hat\sigma^2)r_1^2}{(1+\hat\sigma^2)
(r_1^2+\ell_1^2)}\, 
\nonu
\eea
where $\alpha_0$ was defined previously and $r_1$ and $r_2$ are given
by (\ref{r1r2}).
In this case, since $\hat\sigma$ cannot be zero, there is no undeformed
result. One expects that similar behavior for various $\ell_i, \hat\sigma$
and $t$ can be obtained but it is not too much interested because 
all the quark-monopole potentials are described in the nonzero 
$\hat\sigma$
and we cannot see any phase transition between undeformed theory and
deformed theory.   

There are also two other cases. 
When we compute the $L$, for example, we can take
the first part of $L$ (\ref{EandL}) as an integral $\Delta L$ and the second
part of $L$ (\ref{EandLsecond}) as an integral $L-\Delta L$. Similarly, we
can compute $L$ as a combination of the second part of $L$ in
(\ref{EandL}) for $(1,0)$ string configuration 
and first part of $L$ (\ref{EandLsecond}) for $(0,1)$ string
configuration. Also one can compute the energies $E$ with same regions
of parameter space.   
In each case, there
exist some restrictions on the parameters.
As in second case above, since there is no undeformed solution due to the 
nonzero of $\hat\sigma$, 
we do not know much about any big difference 
between deformed solution and undeformed solution.  


In summary,
we have studied the gravity dual of deformed Coulomb branch of 
${\cal N}=4$ super Yang-Mills theory and applied to 
Wilson loop calculations by minimizing the action for three kinds of
strings on the type IIB supergravity background. 

It would be interesting to consider the nonradial string configuration
\cite{Malda,HSZ} by considering more general solution to 
equations of motion of the action and 
to study how these $\sigma$ deformations change
the confining properties and the phase structure 
of finite temperature \cite{Park}. For the quark-monopole-dyon system 
\cite{DP} where a dyon is added to quark-monopole, 
one can apply $\sigma$ deformation to see how it reflects the 
energy of the system.

\vspace{.7cm}

\centerline{\bf Acknowledgments}

We would like to thank 
Joseph A. Minahan, Justin F. Vazquez-Poritz and Jung-Tay Yee 
for useful discussions. 
This work was supported by grant No.
R01-2006-000-10965-0 from the Basic Research Program of the Korea
Science \& Engineering Foundation. 

\end{document}